\documentclass[prd,preprint,nofootinbib]{revtex4}

\usepackage{graphicx}
\usepackage{amssymb}
\usepackage{amsmath}
\usepackage{color}

\begin{document}
\renewcommand{\thefootnote}{\#\arabic{footnote}}
\newcommand{\rem}[1]{{\bf [#1]}}
\newcommand{\gsim}{ \mathop{}_ {\textstyle \sim}^{\textstyle >} }
\newcommand{\lsim}{ \mathop{}_ {\textstyle \sim}^{\textstyle <} }
\newcommand{\vev}[1]{ \left\langle {#1}  \right\rangle }
\newcommand{\bear}{\begin{array}}  
\newcommand {\eear}{\end{array}}
\newcommand{\bea}{\begin{eqnarray}}   
\newcommand{\eea}{\end{eqnarray}}
\newcommand{\beq}{\begin{equation}}   
\newcommand{\eeq}{\end{equation}}
\newcommand{\bef}{\begin{figure}}  
\newcommand {\eef}{\end{figure}}
\newcommand{\bec}{\begin{center}} 
\newcommand {\eec}{\end{center}}
\newcommand{\non}{\nonumber}  
\newcommand {\eqn}[1]{\beq {#1}\eeq}
\newcommand{\la}{\left\langle}  
\newcommand{\ra}{\right\rangle}
\newcommand{\ds}{\displaystyle}
\newcommand{\red}{\textcolor{red}}
\def\SEC#1{Sec.~\ref{#1}}
\def\FIG#1{Fig.~\ref{#1}}
\def\EQ#1{Eq.~(\ref{#1})}
\def\EQS#1{Eqs.~(\ref{#1})}
\def\lrf#1#2{ \left(\frac{#1}{#2}\right)}
\def\lrfp#1#2#3{ \left(\frac{#1}{#2} \right)^{#3}}
\def\GEV#1{10^{#1}{\rm\,GeV}}
\def\MEV#1{10^{#1}{\rm\,MeV}}
\def\KEV#1{10^{#1}{\rm\,keV}}
\def\REF#1{(\ref{#1})}
\def\lrf#1#2{ \left(\frac{#1}{#2}\right)}
\def\lrfp#1#2#3{ \left(\frac{#1}{#2} \right)^{#3}}
\def\OG#1{ {\cal O}(#1){\rm\,GeV}}

\newcommand{\ah}{A_H}

\title{Decaying Hidden Gauge Boson and the PAMELA and ATIC/PPB-BETS Anomalies}

\author{Chuan-Ren Chen$^{1}$, Mihoko M. Nojiri$^{1,3}$, Fuminobu Takahashi$^{1}$ and T. T. Yanagida$^{1,2}$}

\affiliation{${}^{1}$Institute for the Physics and Mathematics of the Universe,
University of Tokyo, Chiba 277-8568, Japan\\
 ${}^{2}$Department of Physics, University of Tokyo, Tokyo 113-0033,
Japan\\
 ${}^{3}$Theory Group, KEK, and the Graduate University for Advanced Studies (SOKENDAI), 1-1 Oho, Tsukuba, 305-0801, Japan }


\begin{abstract}
We show that the PAMELA anomaly in the positron fraction as well as
the ATIC/PPB-BETS excesses in the $e^- + e^+$ flux are simultaneously
explained in our scenario that a hidden $U(1)_H$ gauge boson
constitutes dark matter of the Universe and decays into the
standard-model particles through a kinetic mixing with an $U(1)_{B-L}$
gauge boson. Interestingly, the $B-L$ charge assignment suppresses an
antiproton flux in consistent with the PAMELA and BESS experiments,
while the hierarchy between the $B-L$ symmetry breaking scale and the
weak scale naturally leads to the right lifetime of $O(10^{26})$
seconds.
\end{abstract}

\preprint{IPMU 08-0092}

\maketitle

The nature of dark matter has been a big mystery in modern cosmology.
Recently, there appeared several exciting observational data on
high-energy cosmic-ray particles, which may be shedding light on this
issue.

The PAMELA data~\cite{Adriani:2008zr} shows that the positron fraction
starts to deviate from the theoretically expected value for secondary
positrons around $10\,$GeV and continues to rise up to $100$\,GeV, and
no drop-off has been observed so far. The excess in the positron
fraction observed by PAMELA strongly indicates that there is an
unidentified primary source of the galactic positrons.  It is natural
to expect that the electron flux may be also modified above
$100$\,GeV, because normally electron-positron pairs are produced 
by such a primary source, and the PAMELA anomaly suggests a
rather hard positron energy spectrum.  Interestingly enough, the ATIC
balloon experiment collaboration~\cite{ATIC-new} measured the total
flux of electrons plus positrons, and has recently released data which
shows a clear excess peaked around $600$\,GeV, in consistent with the
PPB-BETS experiment~\cite{Torii:2008xu}. Since it is difficult to
produce such hard spectrum with a sharp drop-off as observed in the
ATIC/PPB-BETS data by conventional astrophysical sources like pulsars,
the galactic electrons and positrons may be
generated though the annihilation and/or decay of dark matter.
 
We have recently proposed a scenario that a hidden $U(1)_H$ gauge
boson constitutes dark matter of the Universe and decays into the
standard-model (SM) particles through a kinetic mixing with a
$U(1)_{B-L}$ gauge boson~\cite{Chen:2008yi,Chen:2008md}, and it was
shown that our model can explain the PAMELA excess without producing
too many antiprotons, in consistent with the
PAMELA~\cite{Adriani:2008zq} and BESS~\cite{Yamamoto:2008zz}
experiments.\footnote{ See Refs.~\cite{DM-papers} for other dark
  matter models that account for the PAMELA excess. }  In this
letter we show that our model can naturally explain both the PAMELA
and ATIC/PPB-BETS anomalies, for the hidden gauge boson of a mass
about $1.2$\,TeV, while the antiproton flux is still suppressed enough
due to the $B-L$ charge assignment.  Interestingly, our scenario
predicts an excess in the diffuse gamma-ray background peaked around
$100$ GeV, which will be tested soon by the Fermi (formerly
GLAST)~\cite{FGST} satellite in operation.

\vspace{3mm} Let us here briefly review our set-up (see
Ref.~\cite{Chen:2008md} for more details).  We introduce an extra
dimension with two branes at the boundaries.  Suppose that the hidden
gauge sector is on one brane and the SM particles are on the other
brane, which are well separated from each other so that direct interactions
between the two sectors are exponentially suppressed. We assume that a
$U(1)_{B-L}$ gauge field resides in the bulk.  Then the hidden
$U(1)_H$ gauge field can have an unsuppressed gauge kinetic mixing
with the $U(1)_{B-L}$.  We expect that the $U(1)_{B-L}$ gauge symmetry
is broken around the grand unification theory (GUT) scale of about $
10^{15}$ GeV, since the seesaw mechanism~\cite{seesaw} for neutrino
mass generation suggests the right-handed neutrinos of a mass about
$10^{15}$ GeV.  After integrating out the heavy $U(1)_{B-L}$ gauge
boson, the effective couplings between the hidden $U(1)_H$ gauge boson
$A_H$ and the SM particles are induced, which enables $A_H$ to decay
into the SM particles with a extremely long lifetime due to the hierarchy between 
the $B-L$ breaking scale and weak scale.

The low-energy effective interactions between the hidden gauge boson
$\ah$ and the SM fermion $\psi_i$ can be extracted from the
$U(1)_{B-L}$ gauge interactions~\cite{Chen:2008md},
\beq
{\cal L_{\rm int}} \;=\;  
-\lambda\, q_i \frac{m^2}{M^2}  A_H^{\mu} \, \bar{\psi}_i \gamma_\mu \psi_i,
\eeq
where $\lambda$ is a coefficient of the kinetic mixing, $q_i$
denotes the $B-L$ charge of the fermion $\psi_i$, and $m$ and $M$ are
the masses of the hidden gauge boson $A_H$ and the $U(1)_{B-L}$ gauge
boson, respectively.

The partial decay width for the  SM fermion pair is
\bea
\Gamma(\ah \rightarrow \psi_i \bar\psi_i) &\simeq&\lambda^2 \frac{ N_i q_i^2}{12 \pi} \lrfp{m}{M}{4} m,
\label{pGamma}
\eea
where we have neglected the fermion mass, and $N_i$ is the color
factor ($3$ for quarks and $1$ for leptons).  Note that the
coefficient $N_i q_i^2$ is $1/3$ and $1$ for quarks and leptons,
respectively, which results in the suppression of the antiproton
flux. The lifetime $\tau$ is therefore given by
\beq
\tau \;\simeq\; 1\times 10^{26} {\rm \,sec} \left(\sum_i N_i q_i^2\right)^{-1}
\lrfp{\lambda}{0.01}{-2}
 \lrfp{m}{1.2{\rm\, TeV}}{-5} \lrfp{M}{10^{15}{\rm GeV}}{4},
\eeq
where the sum is taken over the SM fermions.  It should be noted that
the branching ratios are not sensitive to the mass of $\ah$ and they
simply reflect the $B-L$ charge assignment, which makes our analysis
very predictive.

Let us now estimate the spectra for the positron fraction,
$(e^-+e^+)$, gamma-ray and antiproton fluxes
based on the decay modes discussed above.  The branching ratios
are $2/39$ and $2/13$ for a quark pair and a charged lepton pair, respectively.
To estimate the spectra of the gamma, positrons and antiprotons, we
use the PYTHIA\ \cite{Sjostrand:2006za} Monte Carlo program. After
cosmic-ray particles are produced during the decay of $A_H$, the
following calculations are straightforward and identical to those
adopted in Ref.~\cite{Chen:2008yi}, and so, we show only the final
results in this letter. For readers who are interested in the details
of the calculations should be referred to Ref.~\cite{Ibarra:2007wg}
and references therein.

\begin{figure}[t]
\includegraphics[scale=0.33]{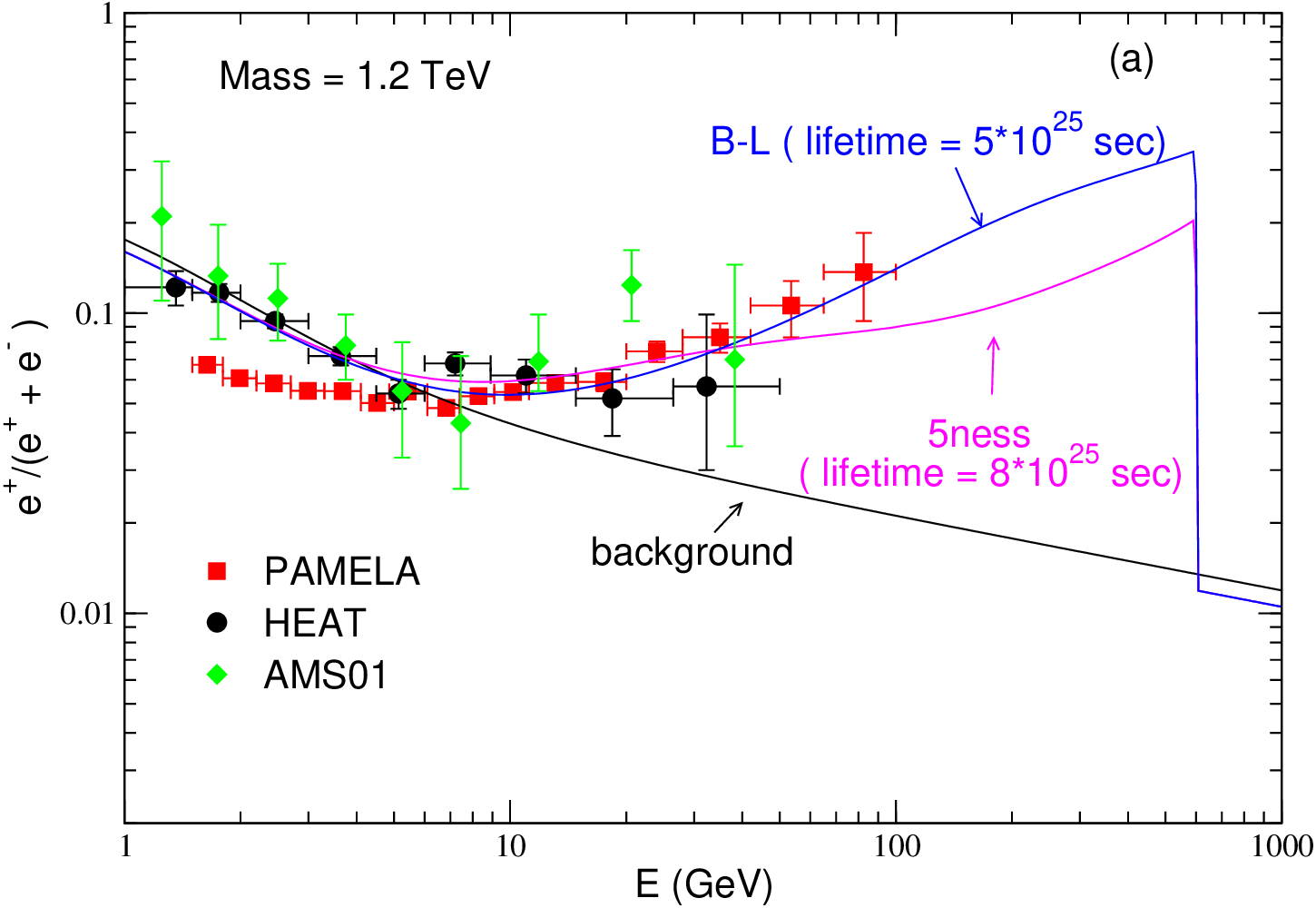}
\includegraphics[scale=0.33]{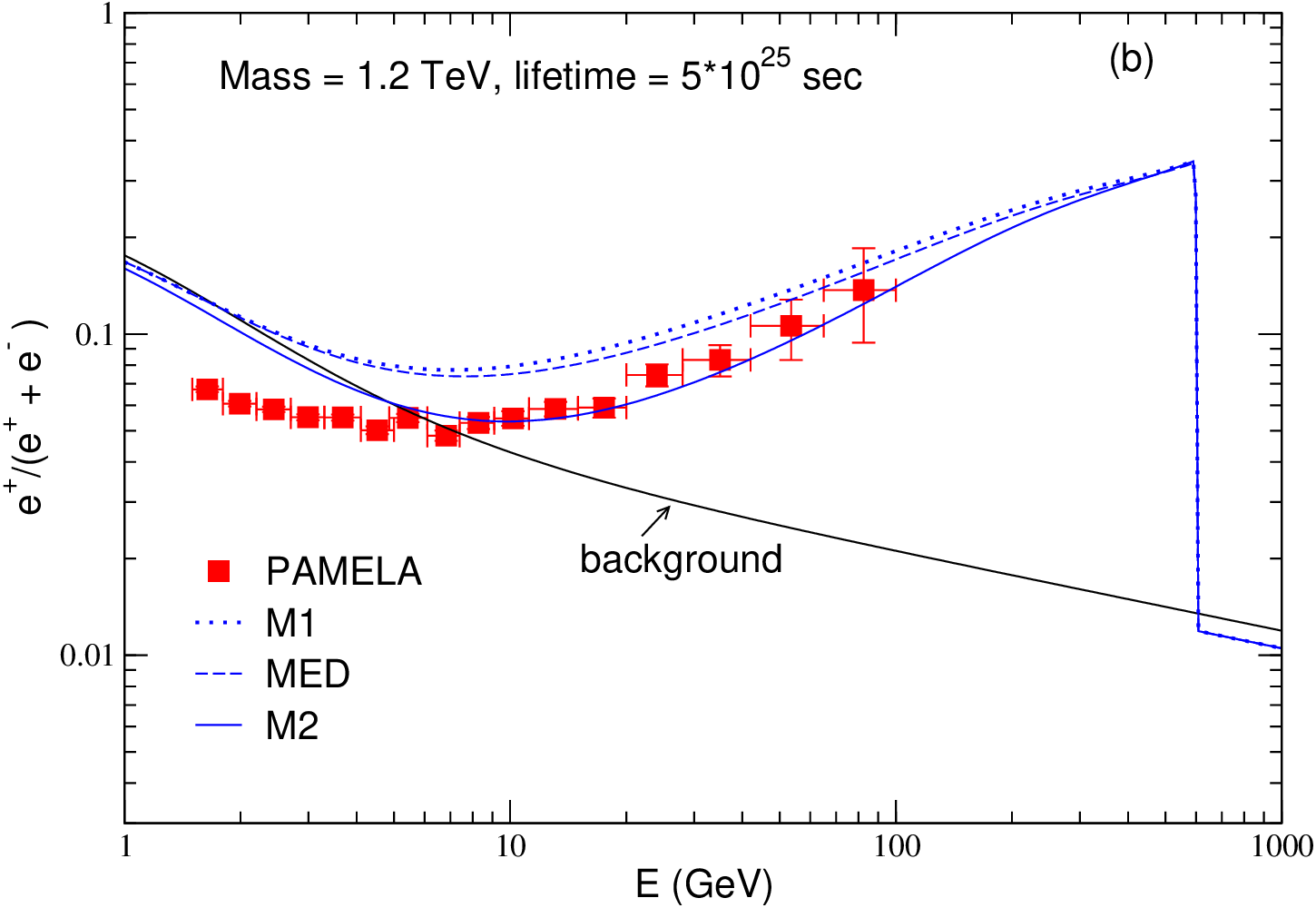}
\caption{(a) The predicted positron fraction from $A_H$ decay via the kinetic mixing with $U(1)_{B-L}$ (blue line) and $U(1)_5$ (magenta line), 
compared with the experimental data~\cite{Barwick:1997ig,Aguilar:2007yf}, including the recent PAMELA results~\cite{Adriani:2008zr}; 
(b) For $U(1)_{B-L}$ case only, using different sets of parameters in solving diffusion equation.}
\label{fig:positron}
\end{figure}

\begin{figure}[t]
\includegraphics[scale=0.326]{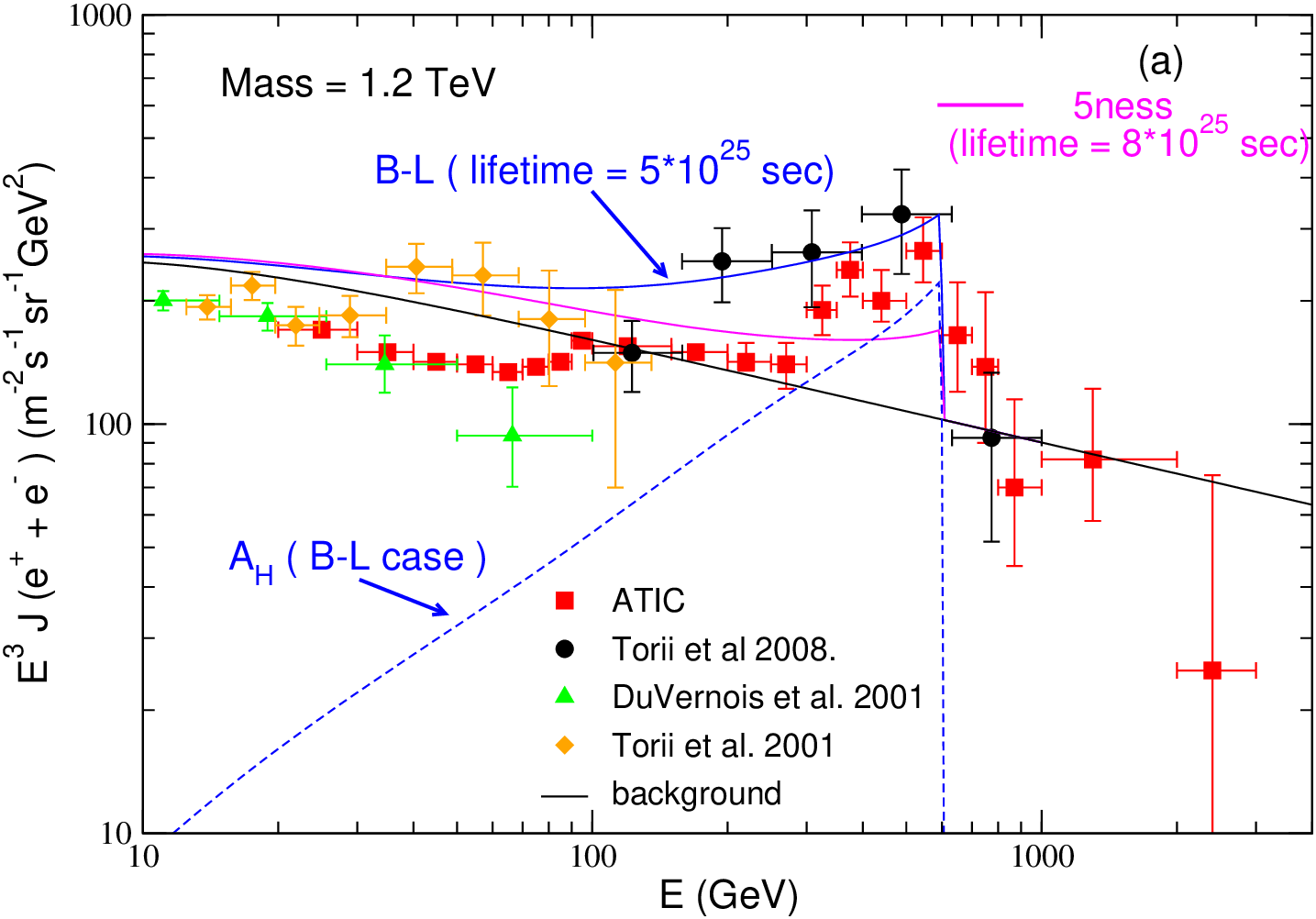}\,\,
\includegraphics[scale=0.326]{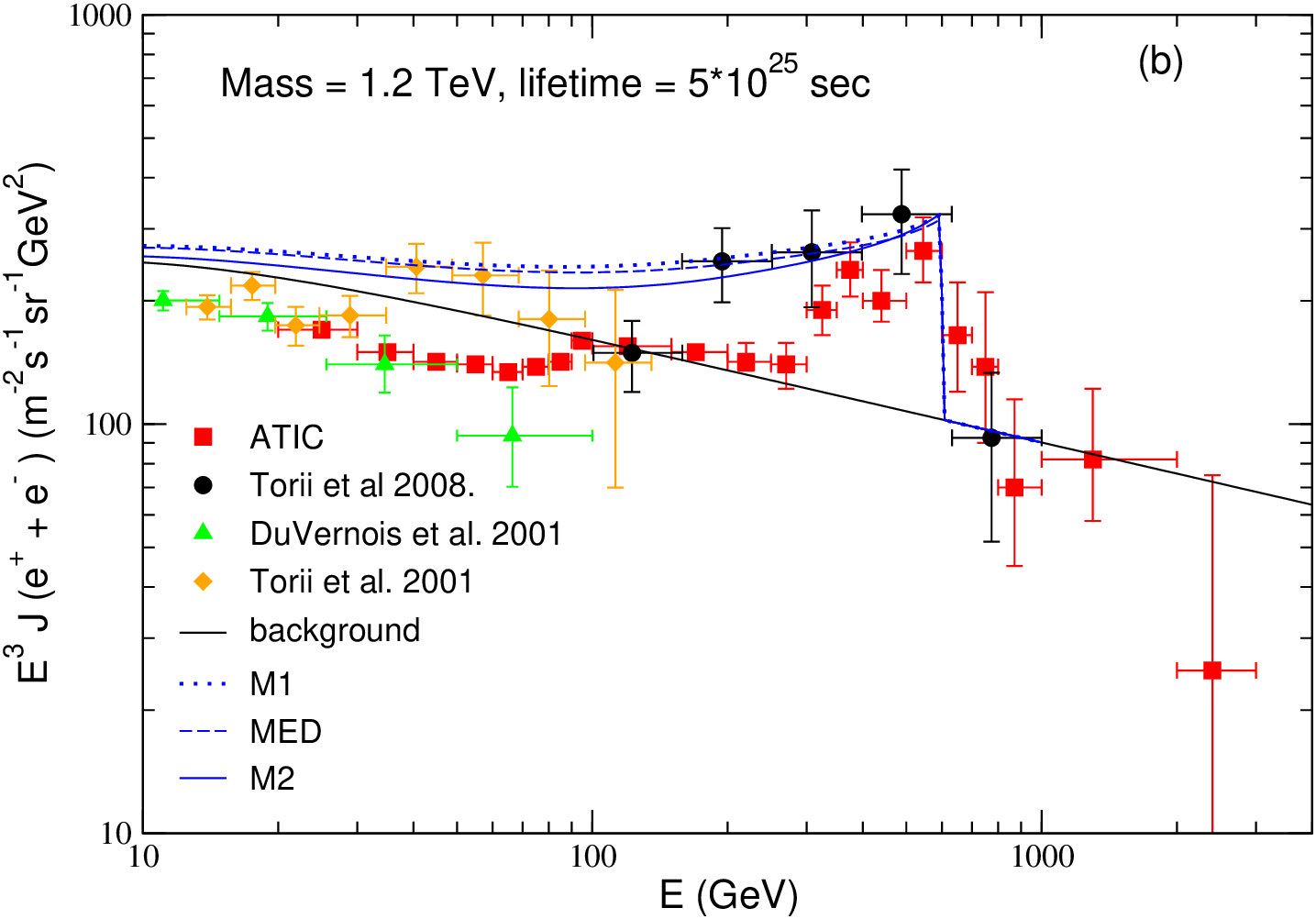}
\caption{(a) The predicted $(e^-+e^+)$ spectrum   from $A_H$ decay via the kinetic mixing with $U(1)_{B-L}$ (blue line)
and $U(1)_5$ (magenta line),  compared with the various observational data~\cite{DuVernois:2001bb,Torii:2001aw}
 including the latest ATIC~\cite{ATIC-new} and 
PPB-BETS~\cite{Torii:2008xu} results. (b) For $U(1)_{B-L}$ case only, using different sets of parameters in solving diffusion equation.}
\label{fig:e+e}
\end{figure}

In our numerical calculations we set $m = 1200$\,GeV and the lifetime
$\tau = 5\times10^{25}$ seconds, and we use the parameter sets that are
consistent with the Boron to Carbon ratio (B/C) and produce the
maximal (M1), medium (MED) and minimal (M2) positron
fluxes~\cite{Ibarra:2007wg}.  In Fig.~\ref{fig:positron} (a), we show the
predicted positron fraction (blue line) together with the recent PAMELA data and
other experiments.  The prediction of our model fits very
well with the PAMELA excess, and increases up to $E= 600$ GeV, a half of the mass of $A_H$.
We also show the $(e^-+e^+)$ spectrum
together with the latest ATIC~\footnote{
The ATIC data points were read from Fig.~3 in Ref.~\cite{ATIC-new}. The background line shown in
Fig.~\ref{fig:e+e} is
slightly different from that adopted in Ref.~\cite{ATIC-new}. Here we have used the same estimate
that we adopted to fit the PAMELA data. Note that, even with a slightly lower background (as in Ref.~\cite{ATIC-new}), 
we can still fit both the PAMELA and ATIC/PPB-BETS excesses by varying the lifetime accordingly.
} and PPB-BETS data in Fig.~\ref{fig:e+e} (a)
(blue line). The contribution from the dark matter decay is shown as the dotted line in
the $B-L$ case, and the characteristic drop-off can be used to infer the mass scale of dark matter.  
Furthermore, the upcoming PAMELA data in higher energy region ($100 \sim 270$\,GeV)
will be able to test our prediction. In Figs.~\ref{fig:positron} (b) and ~\ref{fig:e+e} (b), we 
show the results for different parameters used in solving the diffusion equation. As we can see,
the electron and positron spectrum in the M1 and MED cases are softer
than that in the M2 case.

\begin{figure}[t]
\includegraphics[scale=0.4]{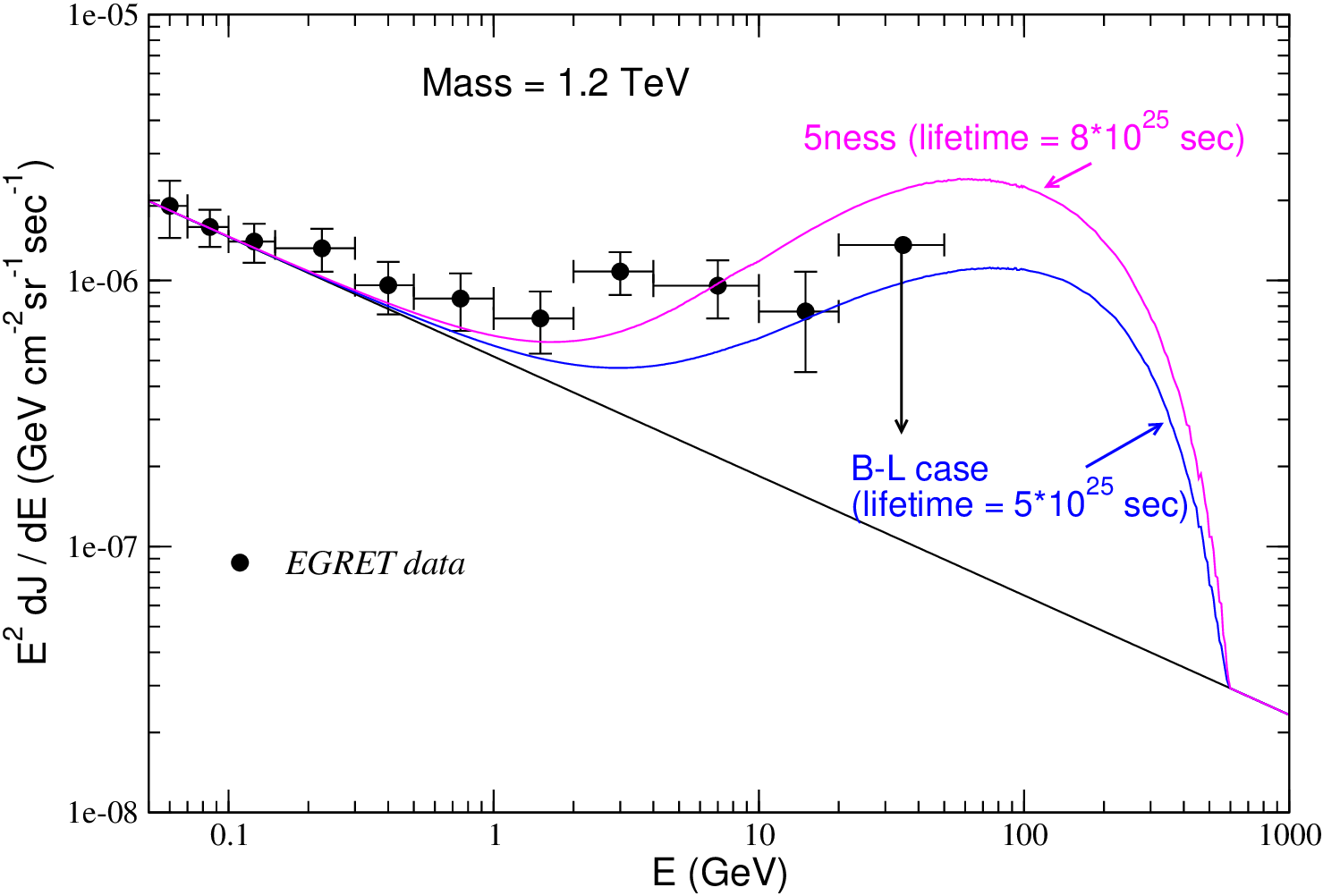}
\caption{The predicted flux of the diffuse gamma-ray background from $A_H$ decay via the kinetic mixing with $U(1)_{B-L}$ (blue line) 
and $U(1)_5$ (magenta line), compared with the EGRET data~\cite{Sreekumar:1997un,Strong:2004ry}}
\label{fig:gamma}
\end{figure}

The gamma-rays are mainly produced by $\pi^{0}$'s generated in the
QCD hadronization process and the decay of $\tau$, since quark pairs as well as a tau lepton
pair are produced from the decay of $A_H$.  In Fig.~\ref{fig:gamma},
we plot the gamma-rays together with the EGRET 
data~\cite{Sreekumar:1997un,Strong:2004ry} (blue line).  The
excesses in the gamma-ray flux are between a few GeV and $600$\,GeV.

\begin{figure}[t]
\includegraphics[scale=0.5]{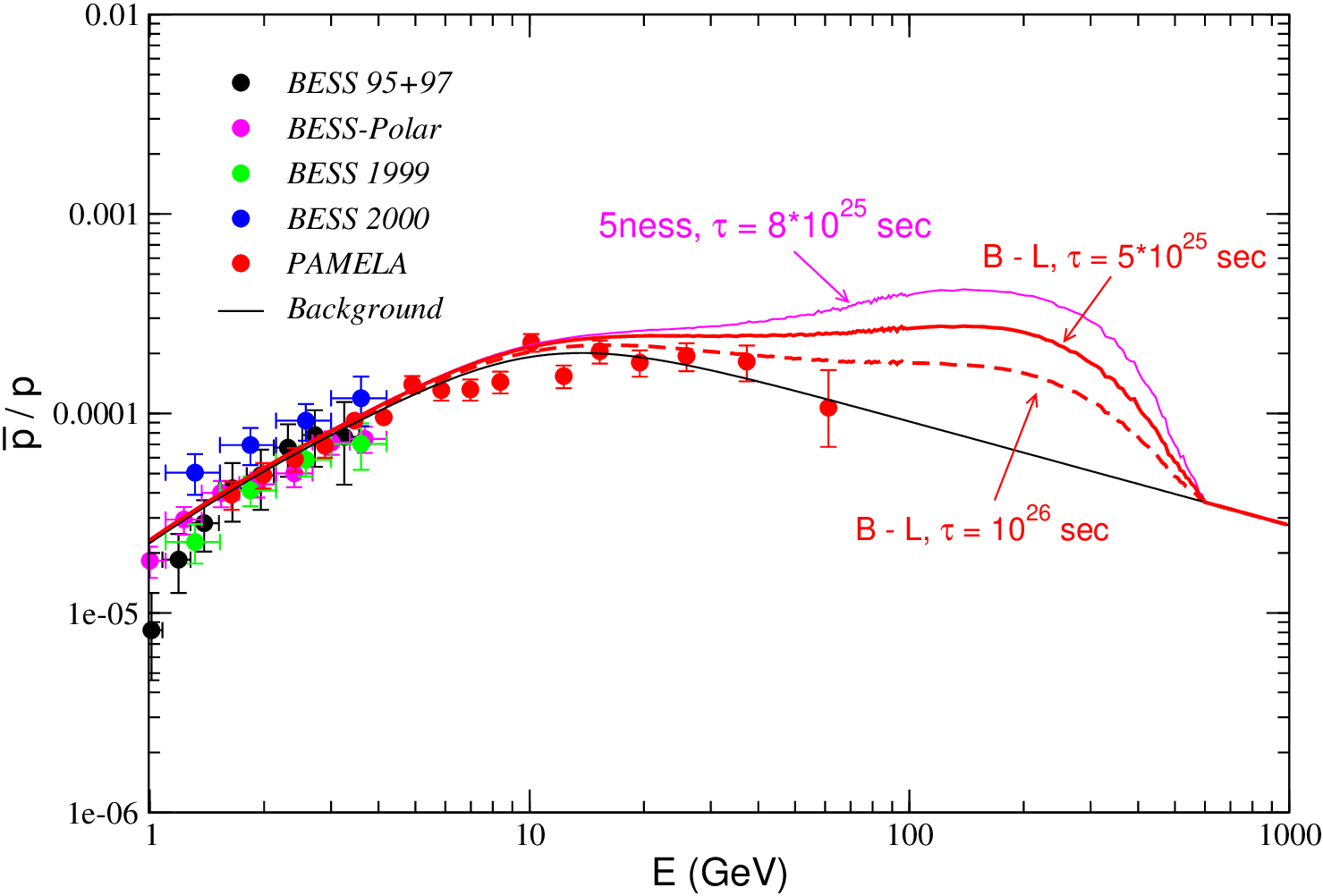}
\caption{The predicted $\bar{p}/p$ ratio and the BESS and PAMELA data.
The red solid (dahsed) lines corresponds to $\tau = 5 \times 10^{25}$\,sec ($10^{26}$\,sec)
for the $B-L$ case, while the purple line represents the fiveness.}
\label{fig:ratio}
\end{figure}

Finally, we show in Fig.~\ref{fig:ratio} the predicted antiproton-to-proton ratio, $\bar{p}/p$, compared with experimental data \cite{Adriani:2008zq,BESS}.
We adopt the MIN diffusion model \cite{Ibarra:2007wg} to calculate the contribution from dark matter. 
For the lifetime $\tau = 5\times10^{25}$ sec, the prediction (solid red line) is consistent with the observational data 
up to $E \lesssim 40$\,GeV, and slightly exceeds the PAMELA data point around $E \simeq 60$\,GeV. This does not necessarily mean
that our model is inconsistent with the PAMELA data.
First, it should be noted that the predicted antiproton flux sensitively depends on the propagation model, 
and the one we adopt is based on several simplifications which has enabled us to solve the diffusion equation analytically. 
Second, the predicted antiproton flux depends on the dark matter profile and model parameters such as mass and lifetime.
For example,  by increasing the mass of dark matter, the place where our prediction starts deviating from the data will be shifted up to higher energy; we may also simply decrease the decay rate. See the red dashed line shown in the Fig.\ref{fig:ratio},
which corresponds to $\tau = 10^{26}$\,sec. In this case, the predicted fluxes of electrons and positrons will be slightly reduced as well.
We may need to adopt different normalization of the primary electrons to have a better fit to the observed data.
However, given relatively large errors of the ATIC/PPB-BETS data, this may not be a severe issue.
In any case, the current PAMELA data in the high energy still does not have  large enough statistics, and
 we expect that the behavior of $\bar{p}/p$ in the higher energy 
will enable us to test or refute the current model in the near future. 

For completeness we consider a case that the $U(1)$ gauge symmetry in
the bulk is $U(1)_5$, so-called ``fiveness", instead of
$U(1)_{B-L}$. The $U(1)_5$ is anomaly free and the charge is
proportional to $4Y-5(B-L)$, where $Y$ is the hypercharge.  We show
the predicted spectra in the case of the $U(1)_5$ as the magenta lines
in Figs.~\ref{fig:positron}, \ref{fig:e+e}, \ref{fig:gamma} and
\ref{fig:ratio}.  Note that the hadronic decay branching ratio, which
is a measure for the antiproton flux, becomes larger, with respect to
that in the $U(1)_{B-L}$ case.

In this letter, we have shown that both the PAMELA excess in the
positron fraction and the ATIC/PPB-BETS anomaly in the electron plus
positron flux are simultaneously explained in our model that the
hidden-gauge-boson dark matter decays into the SM particles via the
kinetic mixing with the $U(1)_{B-L}$ gauge field in the bulk.
Interestingly, our model can naturally avoid the constraint on the
antiproton flux  by PAMELA and BESS experiments due to the smallness of
quark's quantum number under the $U(1)_{B-L}$.  Moreover, our model
predicts an excess in the diffuse gamma-ray background between a few GeV
and $600$\,GeV, which will be tested by the Fermi satellite in
operation. Finally we would like to emphasize that the needed lifetime
of ${\cal O}(10^{26})$ seconds is realized naturally by the hierarchy
between weak scale and the large $B-L$ breaking scale which is about
the GUT scale as suggested by the neutrino masses.

{\it Note added:}
Recently the Fermi LAT collaboration has released the data on the electron plus
positron flux \cite{Abdo:2009zk} , and the ATIC excess was not confirmed. Our model may also 
fit the Fermi data for a different choice of the mass and lifetime,
but detailed analysis is beyond the scope of this letter. We are also aware  that the preliminary result of diffuse gamma-ray from Fermi LAT collaboration  has been presented in several conferences \cite{fermiTalk}. However, we need to wait for the official release of the data in order to
compare it with the prediction of our model.

\begin{acknowledgments} 
F.T. thanks Theory Group of Max-Planck-Institut fur Physik 
for the warm hospitality and support  while part of this work 
was completed.
This work was supported by World Premier International Research Center
Initiative (WPI Initiative), MEXT, Japan. 
\end{acknowledgments}

\end{document}